\documentclass{article}

\newcommand{\hi}{\mathcal{H}}
\newcommand{\his}{\mathcal{H}_{\mathcal{S}}}
\newcommand{\hir}{\mathcal{H}_{\mathcal{R}}}

\newcommand{\Y}{\yen}

\newcommand{\Eff}{\mathcal{E}}

\newcommand{\tr}[1]{\textrm{tr}\left[#1\right]} 
\newcommand{\id}{\mathbbm{1}} 

\newcommand{\R}{\mathcal{R}}
\newcommand*\colvec[3][]{
    \begin{pmatrix}\ifx\relax#1\relax\else#1\\\fi#2\\#3\end{pmatrix}
}
\renewcommand{\S}{\mathcal{S}}

\newcommand{\C}{\mathbb{C}}
\newcommand{\E}{\mathsf{E}}
\newcommand{\F}{\mathsf{F}}
\renewcommand{\R}{\mathcal{R}}

\newcommand{\h}{\hspace{2pt}}
\newcommand{\ha}{\hspace{3pt}}

\newcommand{\pQFrm}{\textbf{pQFrm}_G}
\newcommand{\iQFrm}{\textbf{iQFrm}_G}

\newcommand{\psQFrm}{\textbf{psQFrm}_G}
\newcommand{\QRep}{\textbf{QRep}_G}
\newcommand{\sQRep}{\textbf{sQRep}_G}
\newcommand{\sQInv}{\textbf{sQInv}_G}
\newcommand{\sQEquiv}{\textbf{sQEquiv}_G}

\newcommand{\vNRep}{\textbf{vNRep}_G}
\newcommand{\vNInv}{\textbf{vNInv}_G}

\newcommand{\BG}{\mathbb{B}G}

\newcommand{\hiss}{\hi_{\S'}}
\newcommand{\hirr}{\hi_{\R'}}

\newcommand{\RR}{{\R'}}

\usepackage{physics}
\usepackage{soul}
\usepackage{relsize}
\usepackage{lmodern}
\usepackage{slantsc}
\usepackage{bbm}
\usepackage{graphicx}
\usepackage{amsmath}
\usepackage{bbold}
\usepackage{amsfonts}
\usepackage{amssymb}
\usepackage{amsthm}
\usepackage{amsbsy}
\usepackage{mathrsfs}
\usepackage{varioref}
\usepackage{dsfont}
\usepackage{bm}
\usepackage{color}
\usepackage[nopar]{lipsum}
\usepackage{lipsum}

\usepackage[hyphens]{url}
\usepackage{hyperref}
\usepackage[hyphenbreaks]{breakurl}

\usepackage{upgreek}
\usepackage{booktabs} 
\usepackage{array} 
\usepackage{paralist} 
\usepackage{verbatim}
\edef\restoreparindent{\parindent=\the\parindent\relax}
\usepackage[parfill]{parskip}
\usepackage{tikz-cd}

\usepackage{color}
\usepackage{inputenc}
\usepackage[backend=biber, style=chem-angew, maxbibnames=1, articletitle=true
            ]{biblatex} 
\usepackage[T1]{fontenc}
\newtheorem{theorem}{Theorem}[section] 

\newcommand{\T}{\mathcal{T}}

\newcommand{\hirs}{\hir \otimes \his}

\usepackage{amsmath}
\usepackage[makeroom]{cancel}

\usepackage{enumitem}

\usepackage[lmargin=1.5in, rmargin=1in, tmargin=1in, bmargin=1in]{geometry}
\usepackage{url} 
\usepackage{listings} 
\usepackage{amsmath,amssymb} 
\usepackage{array, booktabs} 
\usepackage{graphicx} 
\usepackage{float} 
\usepackage[export]{adjustbox} 
\usepackage{color} 

\usepackage{comment} 

\usepackage{adjustbox}

\usepackage{hyperref}
\usepackage{cleveref}

\usepackage{pifont}

\begin{document}



\author{Jan G\l{}owacki}

\title{
Relativization is naturally functorial
}
\maketitle

\begin{center}
      \emph{Department of Computer Science, University of Oxford, UK}\\
      \vspace{6pt}\emph{International Center for Theory of Quantum Technologies, University of Gda{\'n}sk, POLAND}\\
      \vspace{6pt}\emph{Basic Research Community for Physics}, Leipzig, GERMANY\\
      \vspace{6pt}\texttt{jan.glowacki.research@gmail.com}\\
\end{center}

\vspace{30pt}

\begin{abstract}

In this note, we provide some categorical perspectives on the relativization construction arising from quantum measurement theory in the presence of symmetries and occupying a central place in the operational approach to quantum reference frames. This construction provides, for any quantum system, a quantum channel from the system's algebra to the invariant algebra on the composite system also encompassing the chosen reference, contingent upon a choice of the pointer observable. These maps are understood as \emph{relativizing} observables on systems upon the specification of a \emph{quantum reference frame}. We begin by extending the construction to systems modelled on \emph{subspaces} of algebras of operators to then define a \emph{functor} taking a pair consisting of a reference frame and a system and assigning to them a subspace of relative operators defined in terms of an image of the corresponding relativization map. When a single frame and equivariant channels are considered, the relativization maps can be understood as a natural transformation. Upon fixing a system, the functor provides a novel kind of frame transformation that we call \emph{external}. Results achieved provide a deeper structural understanding of the framework of interest and point towards its categorification and potential application to local systems of algebraic quantum field theories.

\end{abstract}

\section{Introduction}
Multiple approaches to quantum reference frames are present in the literature, partly sharing conceptual underpinnings but varying significantly in terms of the chosen research strategies. Thus we have an information-theoretic approach (e.g. \cite{bartlett_reference_2007}), the perspective-neutral approach inspired by Dirac quantization of gauge systems (e.g. \cite{de_la_hamette_perspective-neutral_2021}), approaches based on quantum analogues of classical notions of coordinate transformations (e.g. \cite{hamette_quantum_2020, giacomini_einsteins_2023}), and others (e.g. \cite{castro-ruiz_relative_2023,kabel_quantum_2023}). The distinguishing set of underlying ideas of the operational approach to quantum reference frames \cite{loveridge_symmetry_2018,loveridge_relative_2019,loveridge_relativity_2017,glowacki_operational_2023,glowacki_quantum_2023,carette_operational_2023}, on which the present note is meant to provide a categorical perspective, comes from quantum measurement theory. The framework is based on the notion of the \emph{relativization maps} (see below), first introduced in \cite{loveridge_relativity_2017} and generalized to the form studied in this note in \cite{loveridge_symmetry_2018}. Some recent developments were made in \cite{carette_operational_2023} where the suitable notion of an internal frame transformation was introduced and the (ultraweakly closed) images of the relativization maps were identified as frame-relative descriptions of quantum systems. The purpose of this note is to extend the relativization construction to a wider class of systems and frames and capture its functorial and natural properties, providing a better structural understanding of the framework, extending its domain of applicability, and setting the ground for the categorification of its core structure.

The author's motivation for investigating categorical properties of relativization is twofold. Firstly, it sheds light on the structure of the operational quantum reference frames framework, suggesting novel research directions (e.g. external frame transformations, see \ref{exfrtr}). Secondly, the language of Category Theory seems most appropriate for thinking about relationality, which we believe must be a feature of any operationally sound description of physical reality, and thus categorifying the framework may not only generalize it but also, ultimately, provide it with a suitable mathematical foundation.

\section{Preliminaries}

In this section we lay down the functional analytic background necessary to capture the categorical properties of relativization. We begin by introducing the notation for the basic objects of quantum theory and briefly review the necessary notions from the operational approach to quantum reference frames \cite{glowacki_operational_2023,carette_operational_2023}, to then generalize them slightly to the realm of what we call 'semi-quantum' systems.

\subsection{Basics}

\paragraph*{States and effects}
Quantum systems considered are modelled on separable Hilbert spaces $\hi$. The Banach space of bounded operators on such a Hilbert space with the operator norm will be denoted by $B(\hi)$, and the Banach space of trace-class operators with the trace-class norm by $\T(\hi)$. The trace gives a Banach duality between trace-class operators and the bounded ones, i.e, we have $\T(\hi)^* \cong B(\hi)$ in that any continuous linear functional $\T(\hi) \to \C$ is necessarily given by $T \mapsto \tr[TA]$ for some $A \in B(\hi)$. The set of quantum states contains positive trace-class operators on $\hi$ with trace one and will be denoted by $\S(\hi)$, while the set of effects will be written
\[
\Eff(\hi) = \{\F \in B(\hi) \ha | \ha \mathbb{0} \leq \F \leq \id\},
\]
where $A \leq B$ means that $B - A$ is positive, i.e, self-adjoint with non-negative spectrum.

\paragraph*{Topologies}
Besides the norm topologies, we have a useful and operationally motivated dual pair of topologies of point-wise convergence of expectation values on $B(\hi)$ and $\T(\hi)$. Namely, a sequence of bounded operators converges $A_n \to A$ \emph{ultraweakly}\footnote{This topology is also referred to as the $\sigma$-weak or weak$^*$ topology on $B(\hi)$.} iff for any $T \in \T(\hi)$ we have $\tr[T A_n] \to \tr[TA]$; a sequence of trace-class operators converges $T_n \to T$ \emph{operationally} iff for any $A \in B(\hi)$ we have $\tr[T_n A] \to \tr[TA]$. The spaces of states and effects inherit operational and ultraweak topologies from $\T(\hi)$ and $B(\hi)$, respectively. The ultraweak closure operation on the subsets of operator algebras will be denoted by $\{\_\}^{cl}$.

\paragraph*{Quantum channels}

A linear map between the operator algebras $\phi: B(\hi) \to B(\hi')$ is called
\begin{itemize}
    \item \emph{normal} iff it is ultraweakly continuous,
    \item \emph{positive} iff it preserves positivity,
    \item \emph{unital} iff $\phi(\id_\hi)=\id_{\hi'}$.
\end{itemize}
Linear maps that satisfy all the above properties will be referred to as (quantum) \emph{channels}. They preserve effect spaces and have unique predual maps, specified by
\[
    \phi_*: \T(\hi') \to \T(\hi), \h\h 
     \tr[\phi_*(\omega)A] := \tr[\omega \phi(A)],
\]
and preserving state spaces.

\paragraph*{Group representations}
A unitary representation of a locally compact group $G$ on a separable Hilbert space $\hi$ is a group homomorphism $U: G \to B(\hi)^{\rm uni}$, where $B(\hi)^{\rm uni}$ is the group of unitary operators on $\hi$; it gives rise to a (left) action of $G$ on $B(\hi)$ which will be written
\[
    G \times B(\hi) \ni (g,A) \mapsto g.A := U(g) A U(g)^* \in B(\hi).
\]

A channel $\phi: B(\hi) \to B(\hi')$ between quantum systems is \emph{equivariant} iff it intertwines the actions, i.e, for all $A \in B(\hi)$ and $g \in G$ we have
\[
    \phi(g.A)=g.\phi(A).
\]

We will assume all quantum systems considered to be modelled on separable Hilbert spaces $\hi$ equipped with a unitary representation of an arbitrary but fixed locally compact second countable Hausdorff group, giving rise to an ultraweakly continuous action on $B(\hi)$.\footnote{Weak, ultraweak and strong continuity of actions associated to true unitary representations are all equivalent.} The category of quantum systems and channels will be denoted by $\QRep$; the (von Neumann) algebra of invariant operators on $\hi$ will be written $B(\hi)^G$.

\subsection{Quantum reference frames}

We now turn to a brief summary of the basic notions on which the framework of operational quantum frames \cite{glowacki_operational_2023,carette_operational_2023} is built.

\paragraph*{Frame observables} A positive operator-valued measure (POVM) on a locally compact group $G$ is an additive set function
\[
    \E_\R: \BG \to B(\hir),
\]
where $\BG$ denotes the $\sigma$-algebra of Borel subsets, such that $\E_\R(G)=\id$, $\E_\R(\emptyset)= \mathbb{0}$ and for any sequence $\{X_n\}_{n \in \mathbb{N}}$ of \emph{disjoint} Borel subsets we~have
\[
    \E_\R: \bigcup_{n=0}^\infty X_n \mapsto  \sum_{n=0}^\infty \E_\R(X_n).
\]
The elements of the image of $\E_\R$ are called \emph{effects of} $\E_\R$, since indeed they need to belong to the effect space $\Eff(\hir)$.\footnote{Convergence on the right hand side is understood ultraweakly (or, equivalently, weakly since $\Eff(\hi)$ is bounded and hence these topologies agree).} The purpose of a POVM is to assign probability measures over a sample space, here taken to be the group $G$, to quantum states. This is achieved~via the Born rule in the form
\[
    \mu^{\E_\R}_\omega: \BG \ni X \mapsto \tr[\omega \E_\R(X)] \in [0,1], 
\]
where $\omega \in \S(\his)$.

Given an ultraweakly continuous unitary representation $U_\R: G \to B(\hir)^{\rm uni}$, $\E_\R$ is called \emph{covariant} iff for any $X \in \BG$ and $g \in G$ we have
\[
    \E_\R(g.X) = g.\E_\R(X),
\]
where $g.X := \{gh \h | \h h \in X\}$. A covariant POVM on $G$ is understood as a \emph{frame observable} defining a principal\footnote{More general frames can be considered, e.g. with $G$ replaced by a topological space on which $G$ acts transitively.} \emph{quantum reference frame} (or frame/reference for short) $\R := \{U_\R,\E_\R,\hir\}$. Quantum frames are thus modelled by quantum systems with a distinguished observable compatible with the group action. A frame is called \emph{ideal} iff all its effects are projections.

\paragraph*{Relativization}
As has been shown in \cite{loveridge_symmetry_2018}, given a quantum system $\S$ and a principal quantum reference frame $\R$, there is a map $\Y^\R: B(\his) \to B(\hirs)^G$ written as
\begin{equation}
    \Y^\R: B(\his) \ni A_\S \mapsto \int_G d\E_\R(g) \otimes g.A_\S \in B(\hirs)^G,
\end{equation}
called the \emph{relativization map}. It is a normal unital positive linear contraction, and an embedding of von Neumann algebras iff $\R$ is ideal. The space of \emph{relative observables} is defined \cite{glowacki_operational_2023,carette_operational_2023} as the ultraweak closure\footnote{Note that since the invariant subalgebra $B(\hirs)^G$ is a von Neumann algebra, it is ultraweakly closed (weak and ultraweak closures coincide for subalgebras of bounded operators \cite{takesaki_theory_2003}), and hence taking the closure of an invariant subset in $B(\hirs)^G$ is the same as taking it in the whole $B(\hirs)$. In other words, closing the image of $\Y^\R$ ultraweakly in $B(\hirs)$ won't introduce any non-invariant elements, which justifies the inclusion \eqref{def:relop}.} of the image of the relativization map and denoted by
 \begin{equation}\label{def:relop}
    B(\his)^\R := \{\Y^\R(B(\his))\}^{cl} \subset B(\hirs)^G.
 \end{equation}

\section{Semi-quantum systems and frames}

To fully capture the categorical nature of the relativization construction we need to step a little bit outside of the standard setup of quantum mechanics. This is because the spaces of relative operators do not, in general, form operator algebras. However, as ultraweakly closed subspaces thereof, they carry all the structure relevant from the operational perspective.

\paragraph*{Generalized probability theories}

The generalized probability theories (GPTs) that are relevant for this note can be understood as systems slightly more general than the quantum ones. One way to view GPTs is to see them as being \emph{defined} by their \emph{state spaces}. In general, those are taken to be convex subsets of Banach spaces. In the case of quantum theory, these are $\S(\hi) \subset \T(\hi)$. The effects are then understood as maps assigning probabilities to states, so (continuous) affine functionals $e: \S \to [0,1]$, with the space of effects written $\Eff(\S)$. This is compatible with the case of quantum theory when $\F \in \Eff(\hi)$ is identified with the map $\S(\hi) \ni \rho \mapsto \tr[\rho \F] \in [0,1]$.

\paragraph*{Semi-quantum systems}

The generalization of the notion of a quantum system needed in this note is~that of a \emph{semi-quantum system}, for which a state space is an \emph{operational quotient}~\cite{carette_operational_2023}
\[
    \S = \S(\hi)/\hspace{-3pt}\sim_\mathcal{O},
\]
where $\mathcal{O} \subset B(\hi)$ is any subset and $\rho \sim_\mathcal{O} \rho'$ iff $\tr[\rho A] = \tr[\rho' A]$ for all $A \in \mathcal{O}$. Such state spaces are appropriate to consider in a situation when we are given a quantum system but have restricted access to the operators we can apply to the quantum states; some states then become indistinguishable and are therefore identified \cite{glowacki_operational_2023,carette_operational_2023}.\footnote{As an example, an algebra of observables on a chosen \emph{superselection sector} could be chosen as a set of `available observables', or an algebra of local observables of a quantum field. The relation between superselection rules and relative quantities is an unsettled issue (see e.g. \cite{bartlett_reference_2007} and \cite{loveridge_symmetry_2018} for conflicting views).} Due to ultraweak continuity and linearity of the trace, the sets $\mathcal{O}$ and ${\rm span}\{\mathcal{O}\}^{cl}$ give rise to the same semi-quantum state spaces, and the quantum duality extends to this setting in the following sense: semi-quantum state spaces remain Banach spaces (under the quotient norm) and we have a Banach duality \cite{carette_operational_2023}
\[
[\T(\hi)/\hspace{-3pt}\sim_\mathcal{O}]^*\cong {\rm span}\{\mathcal{O}\}^{cl}.
\]
The semi-quantum analogue of the operator algebra $B(\hi)$ is then an \emph{ultraweakly closed} subspace of~such an algebra\footnote{This is similar but different to the notion of an \emph{operator space} \cite{arveson_what_nodate}, which is a norm-closed subspace of $B(\hi)$.} $V \subset B(\hi)$ with the corresponding operational state space given by
\[
\S(V) := \S(\hi)/\hspace{-3pt}\sim_{V}.
\]
We will assume $\id_\hi \in V$, which is the case e.g. whenever the effects of a single POVM are in $V$. When $V$ is an algebra $\mathcal{A}$, being ultraweakly closed it is a von Neumann algebra. In such a case, $\S(\mathcal{A})$ consists precisely of the normal states on $\mathcal{A}$, the Banach space $\T(\hi)/\hspace{-3pt}\sim_\mathcal{A}$ being the (unique) predual.

\paragraph*{Example: relative observables and states}

The primary reason for considering here semi-quantum systems comes from the interpretation of the relative observables to be the only operationally accesible ones \cite{carette_operational_2023,glowacki_operational_2023} after the system and the reference have been specified; the description of the system $B(\his)$ given relative to the reference $\R$ is then given by the semi-quantum system $B(\his)^\R \subset B(\hirs)$. The \emph{relative states} are taken to be classes of states
\[
S(\his)_\R := \S(\hirs)/\hspace{-3pt}\sim_\R,
\]
where $\sim_\R$ is a shorthand for $\sim_{B(\his)^\R}$. There is a bijective correspondence between the relative states and the elements of the image of the predual map $\Y^\R_*$
\[
    \S(\his)^\R := \Y^\R_*(\S(\hirs)) \cong \S(\his)_\R,
\]
which allows for the relative states to be understood as states on the system alone. Depending on the quality of the frame, the set $\S(\his)^\R$ may or may not exhaust $\S(\his)$. The \emph{product-relative} states are those arising from product states on the composite systems, written
\begin{equation}\label{prodrelst}
    \rho^{(\omega)}:= \Y^\R_*(\omega \otimes \rho),
\end{equation}
where $\rho \in \S(\his)$ and $\omega \in \S(\hir)$; see \cite{carette_operational_2023,glowacki_operational_2023} for details and discussions.

\paragraph*{Semi-quantum channels}
The ultraweak topology restricts to subspaces of algebras of bounded operators and allows us to speak of normality, positivity and unitality of linear maps between them in complete analogy to the case of linear maps between algebras of bounded operators. We refer to normal positive unital linear maps with domains and codomains possibly restricted to ultraweakly closed subspaces as \emph{semi-quantum channels}.

If we assume an ultraweakly continuous unitary representation of $G$ on $\hi$, which gives rise to an action on $B(\hi)$ that restricts to an ultraweakly closed subspace $V \subseteq B(\hi)$, i.e, for all $A \in V$ and $g \in G$ we have $g.A \in V$, the subspace $V$ can be considered an object of the category $\sQRep$ of semi-quantum systems with symmetries and (arbitrary) semi-quantum channels.

\paragraph*{Subcategories}

The full subcategory of semi-quantum systems given on von Neumann algebras will be denoted by $\vNRep$. The full subcategory of subspaces of operator algebras on which $G$ acts trivially, objects of which are called \emph{invariant semi-quantum systems}, will be denoted by $\sQInv$. The intersection of the two will be written $\vNInv$.

Notice here that any channel $\phi: V \to V'$ between invariant semi-quantum systems will be equivariant since for any $A \in V$ we have
\[
    \phi(g.A)=\phi(A)=g.\phi(A).
\]

Finally, we will denote by $\sQEquiv \subset \sQRep$ the subcategory of semi-quantum systems and \emph{equivariant} channels, which also turns out to be of interest in the context of relativization.

\paragraph*{Semi-quantum frames}\label{intro:semi-quantum}

Semi-quantum systems can support covariant POVMs, giving rise to the notion of a \emph{semi-quantum reference frame}: one simply needs to choose an ultraweakly closed subspace $V_\R~\hspace{-4pt}\subseteq~\hspace{-4pt}B(\hir)$ in which the effects of the frame observable are all contained; in particular, we always have $\E_\R(G)=\id_{\hir} \in V_\R$. A frame observable can be then understood as a map
\[
    \E_\R: \BG \to V_\R \subseteq B(\hir).
\]
It may then happen that a quantum reference frame could be seen as a semi-quantum one in multiple ways; we view the particular subspace $V_\R \subseteq B(\hir)$ to be part of the definition of a semi-quantum frame. Clearly, given a semi-quantum frame on $V_\R \subseteq B(\hir)$, we can always extend $\E_\R$ with the given inclusion $V_\R \subseteq B(\hir)$ to a POVM in a traditional sense, which allows to use the usual definition of the relativization maps also for semi-quantum frames. The definition of the relative operators then extends trivially to semi-quantum systems $V_\S \subseteq B(\his)$~via
\[
    V_\S^\R := \{\Y^\R(V_\S)\}^{\rm cl} \subset B(\hirs)^G.
\]

\section{Relativization functors}

In this section we show how the assignment $V_\S \mapsto V_\S^\R$, extending $B(\his) \mapsto B(\his)^\R$ from \cite{carette_operational_2023}, can be made \emph{functorial} and discuss its restrictions to different subcategories of systems, of frames; we consider what happens a single frame is fixed to provide relative perspectives on multiple systems, and vice versa when relativization is performed on a fixed system by multiple frames. The first situation sheds light on how relativization interacts with composition of systems, while the second provides a novel notion of~a~frame transformation.

\paragraph*{Category of principal semi-quantum frames}

Let us begin by defining the relevant category of frames. Objects in the category $\psQFrm$ are taken to be principal semi-quantum reference frames, so given by POVMs on $G$ with all the effects contained in semi-quantum systems. As morphism we take the commutative diagrams of the form
    \[
    \begin{tikzcd}
        V_\R \arrow[rr, "\psi"]               &  & V_\RR                  \\
       & \BG \arrow[ul, "\E_\R"] \arrow[ur, "\E_\RR"'] & ,
    \end{tikzcd}
    \]
where $\psi: V_\R \to V_\RR$ is a (necesserily equivariant) semi-quantum channel; thus we write $\psi: \R \to \RR$ whenever the frame observable $\E_\RR$ factorizes through $\E_\R$ via $\psi$, i.e, for all $X \in \BG$ we have
\[
\E_\RR(X) = \psi \circ \E_\R(X).
\]
The full subcategory of principal quantum reference frames, i.e, those for which $V_\R=B(\hir)$, will be written $\pQFrm \subset \psQFrm$. For morphism in $\pQFrm$ to be invertible, $\psi$ needs to be an invertible channel, and thus be given by a unitary $T: \hirr \to \hi_\R$ via $\phi(A) = T A T^*$. Principal quantum reference frames that are isomorphic in $\pQFrm$ are then equivalent in the sense~of~\cite{carette_operational_2023}.

\paragraph*{General definition}

We now provide the relativization functor for semi-quantum systems and principal semi-quantum reference frames. We also discuss some of its conceptually interesting restrictions.

\begin{theorem}
    For any morphism of principal semi-quantum reference frames $\psi: \R \to \RR$ and a~semi-quantum channel $\phi: V_\S \to V_{\S'}$ the map $\Y(\psi,\phi)$, given by the ultraweakly continuous extension~of
    \begin{equation}\label{eq:Ymap}
     \Y(\psi,\phi) : V_\S^\R \ni \int_G d\E_\R(g) \otimes g.A_\S \longmapsto \int_G d(\psi \circ \E_\R)(g) \otimes g.\phi(A_\S) \in V_{\S'}^\RR,
    \end{equation}
    from the dense subset $\Y^\R(V_\S) \subseteq V_\S^\R$ is a well-defined channel between invariant semi-quantum systems extending the assignment $(\R,V_\S) \mapsto V_\S^\R$ to a functor
    \[
        \Y: \psQFrm \times \sQRep \to \sQInv.
    \]
\end{theorem}   

\begin{proof}
    Since $\E_\RR = \phi \circ \E_\R$, clearly
    \begin{equation}\label{eq:yenchannel}
        \Y(\psi,\phi): \Y^\R(A_\S) \mapsto \Y^\RR(\phi(A_\S)),
    \end{equation}
    which in particular assures that the codomain of $\Y(\psi,\phi)$ can indeed be taken to be $V_{\S'}^\RR$. To show linearity on $\Y^\R(V_\S)$, we calculate for arbitrary $\lambda \in \C$ and $A_\S \in V_\S$
    \begin{align*}
        \Y(\psi,\phi) \left(\lambda \Y^\R(A_\S)\right) &= 
        \Y(\psi,\phi) \left(\lambda \int_G d\E_\R(g) \otimes g.A_\S)\right) = 
        \Y(\psi,\phi) \left(\int_G d\E_\R(g) \otimes g.\lambda A_\S)\right)\\ &= 
        \int_G d(\psi \circ \E_\R)(g) \otimes g.\phi(\lambda A_\S)= 
        \lambda \Y(\psi,\phi) \left(\Y^\R(A_\S))\right),
    \end{align*}
    where we have used \eqref{eq:yenchannel} and the linearity of the action $g.\_$ and the channel $\phi$.

    We will now show that $\Y(\psi,\phi)$ is normal on $\Y^\R(V_\S)$. Since $\Y^\R$ is normal and $V_\S$ ultraweakly closed, any sequence $\{B_n\}_{n \in \mathbb{N}} \subset \Y^\R(V_\S)$ converging ultraweakly to $B \in \Y^\R(V_\S)$ can be written as $B_n = \Y^\R(A_n)$ with $A_n \to A$ ultraweakly in $V_\S$ such that $B = \lim_{n \to \infty}B_n = \Y^\R(A)$. Then 
    \[
        \lim_{n \to \infty} \Y(\psi,\phi) (B_n) = \lim_{n \to \infty} \Y^\RR(\phi(A_n)) = \Y^\RR\left(\lim_{n \to \infty} \phi(A_n)\right) = \Y^\RR(\phi(A)) = \Y(\psi,\phi) (B),
    \]
    where we have used \eqref{eq:yenchannel} and normality of $\Y^{\R'}$ and $\phi$. Thus $\Y(\psi,\phi)$ is normal on $\Y^\R(V_\S)$.
    
    Since on any $B(\hi)$ the ultraweak topology is Hausdorff \cite{murphy_c_1990}, and vector space operations are ultraweakly continuous (as follows immediately from the linearity of the trace), ultraweakly closed subspaces of operator algebras are complete Hausdorff topological vector spaces. Existence and uniqueness of the extension $\Y(\psi,\phi)$ to the whole $V_\S^\R$ is then a simple consequence of Theorem 1 in \cite{horvath_topological_1966} (see exercise (U) pg. 139).
    
    Since $\Y^\R$ is positive and $V_\S$ ultraweakly closed, any positive element in $V_\S^\R$ can be ultraweakly approximated by elements $\Y^\R(A_n)$ with $0 \leq A_n \in V_\S$. Since $\Y^\RR$ is also positive and we have \eqref{eq:yenchannel}, the map $\Y(\psi,\phi)$, being normal, also preserves positivity. It is unital as can be seen from the calculation
    \[
        \Y(\psi,\phi)(\id_{\hir \otimes \his}) = \Y(\psi,\phi) \left(\Y^\R(\id_{\his})\right) = \Y^\RR(\phi(\id_{\his})) = \Y^\RR(\id_{\hiss}) = \id_{\hirr \otimes \hiss},
    \]
    where we have used \eqref{eq:yenchannel} and unitality of $\Y^\R$, $\Y^\RR$ and $\phi$. Thus $\Y(\psi,\phi)$ is a well-defined semi-quantum channel for any $\psi: \R \to \R'$ and $\phi: \S \to \S'$.
    
    Regarding functoriality of $\Y$, notice that given $\xi: \R' \to \R''$ and $\lambda: V_{\S'} \to V_{\S''}$ we have
    \begin{align*}
        \Y(\xi,\lambda) \circ \Y(\psi,\phi): \Y^\R(A_\S) \ha \mapsto \ha &\Y(\xi,\lambda) \left(\Y^\RR(\phi(A_\S))\right) =\\
        &\Y^{\R''}(\lambda \circ \phi (A_\S)) = \Y(\xi \circ \psi, \lambda \circ \phi) \left(\Y^\R(A_\S)\right).
    \end{align*}
    Since normal maps compose and the ultraweakly continuous extension is unique, this finishes the proof.
\end{proof}

\paragraph*{Restrictions}

The relativization functor restricts to quantum systems and frames giving
\[
    \Y: \pQFrm \times \vNRep \to \sQInv.
\]
Further, considering only ideal frames allows to describe relativization entirely in the universe of operator algebras. Indeed, the relativization maps for ideal frames, and \emph{only} for such, are multiplicative and define isometric $*$-isomorphisms $\Y^\R:~B(\his) \cong B(\his)^\R$ (e.g. \cite{loveridge_symmetry_2018}). In particular, the image of $\Y^\R$ is then automatically ultraweakly closed, and an invariant von Neumann algebra. This gives a relativization functor of the form
    \[
        \Y: \iQFrm \times \vNRep \to \vNInv,
    \]
where $\iQFrm \subset \pQFrm$ denotes the subcategory of \emph{ideal} quantum frames and \emph{multiplicative} quantum channels. We see it as an interesting feature of relativization that as soon as non-ideal frames are considered, it forces a slight but significant departure from the setup of operator algebras into semi-quantum systems. We also note here, that since the effects of a frame observables of an \emph{ideal} quantum reference frame generate a \emph{commutative} von Neumann algebra, such a frame can be seen as a semi-quantum reference frame on a `classical' system. Genuinely non-classical features of relative descriptions and frame transformations should then only be expected when non-ideal frames are considered. Departure into semi-quantum GPTs then seems unavoidable when non-classicality and relationality, understood in the sense of relativization procedure, are both insisted upon.

\paragraph*{Fixing a frame: naturality of relativization}

As readily seen from the definition of the $\Y$ integral \cite{loveridge_symmetry_2018}, for any equivariant channel $\psi: B(\hir) \to B(\hirr)$ we have
\[
    \psi \otimes \id_\S \int_G d\E_\R(g) \otimes g.A_\S = \int_G d(\psi \circ \E_\R)(g) \otimes g.A_\S.
\]
Further, when we assume $\phi: V_\S \to V_{\S'}$ to also be \emph{equivariant}, we get
\begin{align*}
    \psi \otimes \phi \int_G d\E_\R(g) \otimes g.A_\S &= \int_G d(\psi \circ \E_\R)(g) \otimes \phi(g.A_\S) \\
    &= \int_G d(\psi \circ \E_\R)(g) \otimes g.\phi(A_\S) = \Y(\psi,\phi) \int_G d\E_\R(g) \otimes g.A_\S,
\end{align*}

for all $A_\S \in V_\S$. The relativization functor restricted to
\[
\Y: \psQFrm \times \sQEquiv \to \sQInv
\]
is then simply given by
\begin{equation}\label{eqrel}
    \Y(\psi,\phi) = \psi \otimes \phi: B(V_\S)^\R \to B(V_{\S'})^{\R'}.
\end{equation}

We see that in the context of equivariant channels between systems, the relativization functor is readily \emph{compatible} with the tensor product structure of the operator algebras: the only thing that $\Y$ `does' to a morphism $(\psi,\phi)$ in the product category $\psQFrm \times \sQEquiv$ is to put the channels `parallel' to each other $\psi \otimes \phi$ and adjust the domains and codomains. In this sense, it is compatible with the \emph{monoidal} structure that we have in the underlying category of type I von Neumann algebras. If we now fix the reference $\R$, for any \emph{equivariant} channel $\phi: V_\S \to V_{\S'} \in \sQEquiv$ the relativization functor gives~a~map
\[
    \Y(\id_\R,\phi) = \id_\R \otimes \phi: B(V_\S)^\R \ni \Y^\R(A_\S) \mapsto \Y^\R(\phi(A_\S)) \in B(V_{\S'})^\R.
\]
Moreover, a direct calculation gives
\[
     \Y^\R \circ \phi (A_\S) = \int_G d\E_\R(g) \otimes g.\phi(A_\S) = \int_G d\E_\R(g) \otimes \phi(g.A_\S) = (\id_\R \otimes \phi) \circ \Y^\R (A_\S),
\]
for arbitrary $A_\S \in V_\S$. Since all maps between invariant subspaces are equivariant, the relativization \emph{maps} $\Y^\R$ for a given frame $\R$ then combine to a \emph{natural transformation}
\[
    \Y^\R: {\rm Id} \Rightarrow (B(\hir) \otimes \_)^G,
\]
where both functors are considered as $\sQEquiv \to \sQEquiv$. We see it as a remarkable property of the relativization maps, which then define the relativization functor. Abstracting this construction to the realm of symmetric monoidal categories with objects equipped with representations of a given group is currently being investigated.

\paragraph*{Fixing a system: external frame transformations}\label{exfrtr}

The relativization functor restricted to a fixed semi-quantum system $V_\S \subset B(\his)$ reads
\begin{equation}\label{eq:exfrtr}
    \Y(\psi,\id_\S): V_\S^{\R} \ni \Y^\R(A_\S) \mapsto \Y^{\R'}(A_\S) \in V_\S^{\R'},
\end{equation}
where $\psi:\R \to \R'$ is a frame morphism. We propose to interpret the predual map
\[
\Phi^{\rm ext}_\psi := \Y(\psi,\id_\S)_* : \S(V_\S)^\R \to \S(V_\S)^{\R'}
\]
as a form of an \emph{external} frame transformation.\footnote{We use the word `external' to distinguish from the maps introduced in \cite{carette_operational_2023,glowacki_operational_2023} that are understood as \emph{internal}; here we translate between descriptions of a system given relative to different reference frames, instead of considering all three systems together and picking one of them as a reference to describe the rest.} As easily seen from \eqref{eqrel}, when applied to product-relative states it gives
\[
    \tr[\Phi^{\rm ext}_\psi(\omega' \otimes \rho) \ha \Y^{\R}(A_\S)] = \tr[\psi_*(\omega') \otimes \rho \ha \Y^{\R'}(A_\S)]
\]
for all $\omega' \in \S(V_{\R'})$, $\rho \in \S(V_\S)$ and $A_\S \in V_\S$, which in the notation of \eqref{prodrelst} reads
\[
    \Phi^{\rm ext}_\psi : \rho^{(\omega')} \mapsto \rho^{(\psi_*(\omega'))}.
\]

Thus, transforming a product-relative state along a frame morphism $\psi: \R \to \R'$ simply amounts to applying $\psi_*$ to the frame's state.\footnote{In this sense, it preserves product states and thus the (lack of) \emph{entanglement} between system and (the old and new) reference. Such statements should however be considered with great caution since the states $\omega \otimes \rho$ are only ever evaluated on the relative observables and should be thought of in terms of operational equivalence classes \cite{glowacki_operational_2023,carette_operational_2023} $[\omega \otimes \rho]_\R \in \T(\hirs)/\hspace{-3pt}\sim_\R$.} It is interesting to look at a specific case where $\psi: \R \to \R$ is a frame reorientation, i.e, $\psi = h.\_: V_\R \to V_\R$ for some $h \in G$. We then get
\[
    \Phi^{\rm ext}_{h.\_}: \S(V_\S)^\R \ni \rho^{(\omega)} \mapsto \rho^{(h.\omega)} \in \S(V_\S)^\R,
\]
as expected. We postpone the analysis of properties of external frame transformations when applied to arbitrary relative states to future work.

\section{Summary}
In this note, we have provided a categorical perspective on the relativization construction. We began by generalizing the setup to the realm of \emph{semi-quantum} systems, given by ultraweakly closed subspaces of algebras of bounded operators on separable Hilbert spaces closed under the unitary action of an arbitrary but fixed locally compact second countable Hausdorff group. These kinds of systems generalize quantum systems modelled by representation of groups on von Neumann algebras given on separable Hilbert spaces. Together with normal positive unital maps, they form a category $\sQRep$, with a subcategory $\sQInv$ of \emph{invariant semi-quantum systems}, i.e, subspaces on which the group action is trivial. Principal quantum reference frames, given as covariant POVMs on the group, were likewise generalised to such POVMs with values in semi-quantum systems. Together with equivariant channels between them making such POVMs factorise through one another as arrows, they form the category denoted by $\psQFrm$. We then extended the relativization construction to a functor
    \[
        \Y: \psQFrm \times \sQRep \to \sQInv.
    \]
It is understood as providing descriptions of semi-quantum systems relative to principal semi-quantum reference frames in a way that respects relations between systems and frames. The relative descriptions are $G$-invariant and generically not given in terms of a operator \emph{algebras} but require a step outside into the direction of more general systems, which can be seen by restricting the domain of the relativization functor to
\[
    \Y: \pQFrm \times \vNRep \to \sQInv.
\]

However, when restricted to \emph{ideal} frames and multiplicative channels between them, the realm of von Neumann algebras is enough to capture functoriality of relativization as the functor restricts~to
\[
    \Y: \iQFrm \times \vNRep \to \vNInv.
\]

Upon fixing a semi-quantum system, the relativization functor provides a map between its descriptions relative to different frames, and can thus be interpreted as providing an \emph{external} quantum reference frame transformations. We established immediate properties of such transformations when applied to product-relative states.

Further, we noticed that when the frame is fixed and we restrict to the subcategory of equivariant channels between systems, the relativization functor simplifies significantly and assigns to a pair of morphisms $(\psi,\phi): (\R,\S) \to (\R',\S')$ the tensor product channel with adjusted domain and codomain, i.e,
\[
\Y(\psi,\phi) = \psi \otimes \phi: B(V_\S)^\R \to B(V_{\S'})^{\R'},
\]
showing the compatibility of the relativization with the underlying monoidal structure. Following this line of thought we showed that the relativization maps form a natural transformation
\[
\Y^\R: {\rm Id} \Rightarrow (\R \otimes \_)^G,
\]
where both functors are considered as $\sQEquiv \to \sQEquiv$. In future work, we plan to further investigate the properties of general external frame transformations, and to provide a purely categorical characterization of the discovered structures.

\paragraph*{Acknowledgments}

I would like to thank here Nesta van der Shaaf, Titouan Carrete, Leon Loveridge,
Chris Heunen, John Selby and Aleks Kissinger for their support and inspiring conversations.

I also acknowledge the funding received via NCN through the OPUS grant nr.~$2017/27/$B/
ST$2/02959$ and support by the Digital Horizon Europe project FoQaCiA, Foundations of quantum computational advantage, GA No. 101070558, funded by the European Union, NSERC (Canada), and UKRI~(UK).

This publication was made possible through the support of the ID\# 62312 grant from the John Templeton Foundation, as part of the \href{https://www.templeton.org/grant/the-quantuminformation-structure-ofspacetime-qiss-second-phase}{‘The Quantum Information Structure of Spacetime’ Project (QISS)}. The opinions expressed in this project/publication are those of the author(s) and do not necessarily reflect the views of the John Templeton Foundation.

\printbibliography[title={References}]

\end{document}